 \documentclass[aps,pra,superscriptaddress,amsmath,amssymb,preprintnumbers,twocolumn,floatfix,showpacs,showkeys,10pt]{revtex4-1}
 \usepackage{amssymb} \usepackage{epsfig}
 \begin{document}
  \title{Memory effects teleportation of quantum Fisher information under decoherence}

\author{You-neng Guo}
\email{guoxuyan2007@163.com}
\affiliation{Department of Electronic and Communication Engineering, Changsha University, Changsha, Hunan
410022, People's Republic of China}
\affiliation{Interdisciplinary Center for Quantum Information, National University of Defense Technology, Changsha 410073, People's Republic of China}
\author{Ke Zeng}
\email{corresponding author:zk_92@126.com}
\affiliation{Department of Electronic and Communication Engineering, Changsha University, Changsha, Hunan
410022, People's Republic of China}
\author{Ping-xing Chen}
\email{corresponding author:pxchen@nudt.edu.cn}
\affiliation{Interdisciplinary Center for Quantum Information, National University of Defense Technology, Changsha 410073, People's Republic of China}

\begin{abstract}
We have investigated how memory effects on the teleportation of quantum Fisher information(QFI) for a single qubit system using a class of X-states as resources influenced by decoherence channels with memory, including amplitude damping, phase-damping and depolarizing channels. Resort to the definition of QFI, we first derive the explicit analytical results of teleportation of QFI with respect to weight parameter $\theta$ and phase
parameter $\phi$ under the decoherence channels. Component percentages, the teleportation of QFI for a two-qubit entanglement system has also been addressed. The remarkable similarities and differences among these two situations are also analyzed in detail and some significant results are presented.
\end{abstract}


 \maketitle
\section{Introduction}

Quantum teleportation firstly proposed by Bennett et al~\cite{Bennett}, becomes one of the most fascinating protocols in quantum communication and quantum computation networks. With the help of priori quantum entanglement shared by the sender and receiver, teleporting an unknown quantum state from one partner to another without transferring the physical carrier of the state is feasible by performing local quantum operations and classical communication. In the past few decades, quantum teleportation has attracted considerable attentions and developed various approaches not only in theories but also in experiments~\cite{Pfaff,Espoukeh,Wang,Takesue,Joo,Mohammadi,Mirmasoudi,Kogias,Jun}. Particularly, it has recently been realized over 1000-kilometer free-space channel quantum teleportation in the experiments~\cite{Ren}.

From the perspective of transmission information security, people however, pay more attention to the information of a particular parameter physically encoded in the teleported state rather than the whole quantum state itself in the realistic process. Transferring the whole quantum state directly without encoding is more susceptible to the effect of environmental noise, which leads to the occurrence of bit error codes so that the transformation information is distorted seriously. Meanwhile, the information teleportation that carried by a physical parameter makes ensure the reliability of communication and is not easy to eavesdrop.
On the other hand, similarly to conventional quantum state teleportation where the quality of teleportation is quantified by fidelity, the teleportation of information of specific parameters encoded
in quantum states can be quantified by quantum Fisher information (QFI)~\cite{Helstrom} which is the credibility of particular information teleportation, describing how accurate a parameter encoded into the quantum state of the system is estimated. In fact, the idea of considering the QFI teleportation first appeared in the Ref.~\cite{Lu} where the authors used the QFI in characterizing the information flow of open quantum systems, and was subsequently generalized to investigate the dynamics of QFI under various noisy environments as well as QFI teleportation~\cite{Song,Xiao,Yao0,Jafarzadeh,Metwally,Metwally1,Ying}.
As a further step along this line, we address how memory effects on the teleportation of QFI in quantum channels with memory. Different from the non-Markovian memory effects in open quantum systems theory where memory effects arise dynamically during the time evolution of quantum systems~\cite{Breuer0,Fanchini}, quantum channels with memory are characterized by the
existence of correlations between successive applications of the channel on a sequence of quantum
systems in quantum information theory~\cite{Caruso,Nigum,Kretschmann,Yeo1}.

Based on the standard teleportation protocol proposed in Refs.~\cite{Bowen,Bowen1}, our work aims to concern only the transformation of information of specific parameters encoded
in the quantum state but not the whole quantum state itself. By using the QFI definition, we have explored the teleportation of QFI for a class
of X-states as resources under quantum channels with memory. We first derive the explicit analytical results of teleportation of QFI with respect to weight parameter $\theta$ and phase
parameter $\phi$ suffering from decoherence. Our analytical results show that the precision of estimating parameters, are strongly determined by the parameters of initial state $c_{i}$, teleported sate $\theta$ and $\phi$, as well as noisy channels $D$ and $\mu$ during the teleportation process. Meanwhile, we also consider the teleportation of a two-qubit entanglement sate subjected to decoherence, and compare the situations where the teleportation of QFI for a single- and two-qubit state are to teleport. The remarkable similarities and differences among these two situations under noisy channels are also analyzed in detail and some significant results are presented. The results show that memory effects can improve the precision of estimating parameters.

The layout is as follows: In Sec. \textrm{II}, we illustrate noise channels and teleportation protocol. In Sec. \textrm{III}, we devote to examining the QFI teleportation of a single qubit state in different types of noisy channels with memory. In Sec. \textrm{IV}, teleportation of QFI of a two-qubit state is investigated. Finally, we give the conclusion  in Sec. \textrm{V}.

\section{Noise model and teleportation}  %%% ½Ú±êÌâ 2
Quantum channels model noise processes that occur in quantum systems due to the interaction with their environments~\cite{Breuer}. Mathematically, quantum channel $\varepsilon$ is a completely positive and trace preserving map of a quantum system from an initial state $\rho$ to the final state $\varepsilon(\rho)=\sum_{i}E_{i}\rho E_{i}^{\dagger}$, where $E_{i}=\sqrt{P_{i_{1}...i_{N}}}A_{i}$ are the Kraus operators of the channel which satisfies the completeness relationship, and $\sum_{i}P_{i_{1}...i_{N}}=1$. $P_{i_{1}...i_{N}}$ can be interpreted as the probability that a random sequence of operations are applied to the sequence of
$N$ qubits transmitted through the channel. For a memoryless channel, where environmental correlation time is far smaller than the time between consecutive
uses, so that at each channel use the environment back action can be negligible. Namely, the system undertakes the same quantum channel $\varepsilon$, in which independent noise acts on each use. Suppose $N$ times uses of this channel, we have that $\varepsilon_{N}=\varepsilon_{}^{\otimes N}$. The Kraus operations $A_{i}$ are independent, and $P_{i_{1}...i_{N}}=P_{i_{1}}P_{i_{2}}...P_{i_{N}}$. In real physical quantum transmission channels, however, it is common to have correlated noise acting on consecutive uses. These kinds of channels are called memory channels that can happen when environmental correlation time is much larger than the time between consecutive uses, so that the channel acts dependently on each channel input $\varepsilon_{N} \neq \varepsilon_{}^{\otimes N}$ and $P_{i_{1}...i_{N}}=P_{i_{1}}P_{i_{2}|i_{1}}...P_{i_{N}|i_{N-1}}$, here $P_{i_{N}|i_{N-1}}$ is the conditional probability of an operation. particularly, for two consecutive uses of a channel with partial memory, the Kraus operators are taken form as
\begin{equation}
E_{i,j}=\sqrt{P_{i}[(1-\mu)P_{j}+\mu\delta_{i,j}]}A_{i}\otimes A_{j}
\label{eq2}
\end{equation} where $0\leq \mu \leq 1$ is the memory coefficient of channel. Physically the parameter $\mu$ is determined by the relaxation
time of the channel when a qubit passes through it.

Following, we consider the noisy channels with memory for two consecutive uses, including amplitude damping, phase damping and depolarizing channels (see Table I in Appendix A). For phase damping and depolarizing channels with uncorrelated case, the Kraus operators are taken forms as $E_{i,j}^{u}=\sqrt{P_{i}}\sqrt{P_{j}}\sigma_{i}\otimes\sigma_{j}$, while  $E_{k,k}^{c}=\sqrt{P_{k}}\sigma_{k}\otimes\sigma_{k}$ for these channels with correlated noise with $i,j,k=0,1,2,3$. $\sigma_{0}=I$ is identity matrix and $\sigma_{1(2,3)}$ are Pauli operators. However, for amplitude damping channel with correlated case, the Kraus operators can not be constructed in a similar manner. According to Ref.~\cite{Yeo1,Macchiavello}, the correlated operators $E_{k,k}^{c}$ are determined by solving the correlated Lindblad equation. Mathematically, for any initial state $\rho$, the finial output state under these channels with memory in terms of the Kraus operator approach, is determined by
\begin{equation}
\varepsilon(\rho)=(1-\mu)\sum_{i,j}E_{i,j}^{u}\rho E_{i,j}^{u\dagger}+\mu\sum_{k}E_{k,k}^{c}\rho E_{k,k}^{c\dagger}
\label{eq3}
\end{equation}
The above expression means that the same operation applied to both qubits with probability $\mu$ are correlated while with probability $1-\mu$ are uncorrelated.

Assume the initial state resource setting of the communication channel between the partners is prepared in the following two-qubit X-state
\begin{equation}
\rho=\frac{1}{4}(\sigma_{0}\otimes \sigma_{0}+\Sigma_{i=1}^{3}c_{i}\sigma_{i}\otimes \sigma_{i})
\label{eq4}
\end{equation}
with $0\leq |c_{i}|\leq 1$. For $|c_{1}|=|c_{2}|=|c_{3}|=c$, Eq.~\eqref{eq4} reduces  to Werner states. Especially, $|c_{1}|=|c_{2}|=|c_{3}|=1$ is corresponding to the maximal entangled states. Submitting  Eq.~\eqref{eq4} into Eq.~\eqref{eq3}, it is straightforward to obtain the output states for different noisy channels with memory, and have forms

\begin{equation}\varepsilon(\rho)= \left(
\begin{array}{ c c c c l r }
\rho_{11} & 0 & 0 & \rho_{14} \\
0 & \rho_{22} & \rho_{23} & 0 \\
0 & \rho_{32} & \rho_{33} & 0 \\
\rho_{41} & 0 & 0 & \rho_{44} \\
\end{array}
\right)
\label{eq5}
\end{equation}

In order to perform quantum teleportation protocol, we assume that the sender and the receiver shared a mixed state given in Eq.~\eqref{eq5} as quantum channel resource, to teleport an arbitrary unknown qubit pure state $|\psi\rangle_{A}$ from one to the other, where $|\psi\rangle_{A}$ is taken form
\begin{equation}
|\psi\rangle_{A}=\cos\frac{\theta}{2}|0\rangle+e^{i\phi}\sin\frac{\theta}{2}|1\rangle
\label{eq6}
\end{equation}
with $0 \leq \theta \leq \pi$ and $ 0 \leq \phi < 2\pi$. In fact, teleportation protocol which is using a mixed state as quantum channel resource is tantamount to a noisy channel, has been proved in
the standard teleportation protocol $\mathcal{T}_{0}$ by Bowen and Bose~\cite{Bowen}. According to their the results, an input state is destroyed and its replica (output) state appears at remote
place after applying a local measurement and unitary
transformation in the form of linear operators:
\begin{equation}
\Lambda_{\mathcal{T}_{0}}(|\psi\rangle\langle\psi|)=\sum_{i=0}^{3}\langle\Psi_{Bell}^{i}|\varepsilon(\rho)|\Psi_{Bell}^{i}\rangle\sigma_{i}(|\psi\rangle_{A}\langle\psi|)\sigma_{i}
\label{eq7}
\end{equation}
where $|\Psi_{Bell}^{i}\rangle$ are the four maximally Bell entangled states: $|\Psi_{Bell}^{0}\rangle=\frac{1}{\sqrt{2}}(|01\rangle-|10\rangle)$,
$|\Psi_{Bell}^{1}\rangle=\frac{1}{\sqrt{2}}(|00\rangle-|11\rangle)$, $|\Psi_{Bell}^{2}\rangle=\frac{1}{\sqrt{2}}(|00\rangle+|11\rangle)$ and $|\Psi_{Bell}^{3}\rangle=\frac{1}{\sqrt{2}}(|01\rangle+|10\rangle)$.

To quantify the amount of teleported information,  we are interested in the precision of estimating the parameters $\theta$ and $\phi$ which are encoded in the teleported state(or the input state) by means of QFI.

\section{QFI teleportation}
QFI as a crucial concept, has been extensively investigated in parameter estimation theory. The parameter estimation precision is  bounded by the quantum Cram\'{e}r-Rao inequality\cite{Helstrom,Holevo} i.e., $\Delta\phi\geq1/\sqrt{n \mathcal {F}_{\phi}}$, where $n$ is the number of experiments and $\mathcal {F}_{\phi}$ denotes the QFI of the parameter $\phi$. According to the symmetric logarithmic derivative operator $\mathcal {L}_{\phi}$ , the QFI of $\phi$ is defined as $\mathcal {F}_{\phi}=\textrm{Tr}[(\partial_{\phi} \rho_{\phi})\mathcal {L}_{\phi}]$, where
$\partial_{\phi}\rho_{\phi}=(\mathcal
{L}_{\phi}\rho_{\phi}+\rho_{\phi}\mathcal {L}_{\phi})/2$
with $\partial_{\phi}=\partial/\partial\phi$. Making use of the spectrum decomposition $\rho_{\phi}=\Sigma_{i=1}^{M}|\lambda_{i}|\psi_{i}\rangle\langle\psi_{i}|$, the expression of QFI can be rewritten as\cite{Knysh,Liu}
\begin{eqnarray}\label{eq2}
\mathcal{F}_{\phi}&=&\sum_{i=1}^{M}\frac{(\partial_{\phi}\lambda_{i})^2}{\lambda_{i}}+2\sum_{i\neq j}^{M}\frac{(\lambda_{i}-\lambda_{j})^2}{\lambda_{i}+\lambda_{j}}|\langle\psi_{i}\partial_{\phi}\psi_{j}\rangle|^2
\end{eqnarray}
Note that $\mathcal{F}_{\phi}=\langle\psi_{i}|\partial_{\phi}\psi_{i}\rangle-|\langle\psi_{i}\partial_{\phi}\psi_{j}\rangle|^2$ is corresponding to the pure state $|\Psi\rangle$. However, for any single qubit state $\rho$, that can be written the form as $\rho = \frac{1}{2}(1+\vec{r}\cdot\hat{\sigma})$ in the Bloch sphere representation, where $\vec{r}=(r_{x}, r_{y}, r_{z})$ is the real Bloch vector
and $\hat{\sigma}=(\hat{\sigma}_{1}, \hat{\sigma}_{2},\hat{\sigma}_{3})$ denotes the Pauli matrices. In terms of the Bloch sphere representation, ~\eqref{eq2} can be rewritten as\cite{Zhong}
\begin{eqnarray}
\mathcal{F}_{\phi}=\left\{\begin{array}{cc}
|\partial_{\phi} \vec{r}|^{2}+\frac{(\vec{r}\cdot\partial_{\phi}\vec{r})^2}{1-|\vec{r}|^{2}}, & \mbox{ if } \, |\vec{r}|<1,\\
|\partial_{\phi}\vec{r}|^{2}, & \mbox{ if } \, |\vec{r}|=1.
\end{array}\right.
\label{eq1}
\end{eqnarray}

Having introduced the definition of QFI, it is straightforward to calculate the QFI of the input state with respect to $\theta$ and $\phi$, $\mathcal{F}_{\theta}=1$ and $\mathcal{F}_{\phi}=\sin^2\theta$, respectively. Obviously, the QFI with respect to $\theta$ ($\mathcal{F}_{\theta}$) is constant and independent of both parameters $\theta$ and $\phi$, while the QFI with respect to $\phi$ ($\mathcal{F}_{\phi}$) is related to $\theta$ and symmetric with respect to $\theta=\pi/2$. In particular, $\mathcal{F}_{\phi}$ has a maximum value 1 at $\theta=\pi/2$.

In the following sections, we are motivated to discuss the precision of estimating parameters both $\theta$ and $\phi$ by means of QFI which are influenced by amplitude damping, phase damping and depolarizing channels with memory, respectively.

\subsection{Amplitude damping channel with memory}
Amplitude damping channel which is used to characterize spontaneous emission represents the dissipative interaction between the system and
the environment. As described above, the partners shared the state~\eqref{eq5} as a communication channel to teleport an unknown state from
one to the other using the Bowen and Bose's protocol. Combination Eqs.~\eqref{eq3} with ~\eqref{eq7}, the teleported final state by means of its Bloch vector
\begin{eqnarray}
r_{x}&=&\frac{1}{2}(Bc_{1}-Ac_{2})\cos\phi\sin\theta\nonumber \\
r_{y}&=&\frac{1}{2}(Bc_{2}-Ac_{1})\sin\phi\sin\theta\nonumber \\
r_{z}&=&\frac{1}{2}M\cos\theta
\end{eqnarray}
with $A=(1-\sqrt{1-D})\mu$, $B=A-2[1-D(1-\mu)]$ and $M=2c_{3}[(2-D)D(1-\mu)-1]-2D^2(1-\mu)$.
Using Eq.~\eqref{eq1}, the teleported QFI with respect to $\theta$

\begin{widetext}
\begin{eqnarray}\label{Eq8}
\mathcal{F}_{\theta}&=&\frac{1}{4}[(Bc_{1}-Ac_{2})^2\cos^2\theta\cos^2\phi+(Bc_{2}-Ac_{1})^2\cos^2\theta\sin^2\phi+M^2\sin^2\theta]\nonumber \\
&+&\frac{\cos^2\theta\sin^2\theta[(Bc_{1}-Ac_{2})^2\cos^2\phi+(Bc_{2}-Ac_{1})^2\sin^2\phi-M^2]^2}{4-(Bc_{1}-Ac_{2})^2\cos^2\phi\sin^2\theta-(Bc_{2}-Ac_{1})^2\sin^2\theta\sin^2\phi-M^2\cos^2\theta}
\end{eqnarray}
and the teleported QFI with respect to $\phi$
\begin{eqnarray}\label{Eq9}
\mathcal{F}_{\phi}&=&\frac{1}{4}[(Bc_{2}-Ac_{1})^2\sin^2\theta\cos^2\phi+(Ac_{2}-Bc_{1})^2\sin^2\theta\sin^2\phi]\nonumber \\
&+&\frac{\cos^2\phi\sin^2\phi\sin^4\theta[(A^2-B^2)(c_{1}^2-c_{2}^2)]^2}{4[4-(Bc_{1}-Ac_{2})^2\cos^2\phi\sin^2\theta-(Bc_{2}-Ac_{1})^2\sin^2\theta\sin^2\phi-M^2\cos^2\theta]}
\end{eqnarray}
\end{widetext}
From the above equations, one can find that the teleportation of QFI $\mathcal{F}_{\theta}$($\mathcal{F}_{\phi}$) are determined by the initial parameters $c_{i}$, teleported sate parameters $\theta$ and $\phi$, as well as noisy channel parameters $D$ and $\mu$ during the teleportation process. Obviously, for any of maximal Bell entangled states as quantum resources(e.g. $|c_{1}|=|c_{2}|=|c_{3}|=1$) subjected to decoherence, the QFI of the teleportted state with respect to $\theta$ and $\phi$, reduce to $\mathcal{F}_{\theta}=1$ and $\mathcal{F}_{\phi}=\sin^2\theta$ in the limit $\mu\rightarrow 1$, respectively.

In order to estimate the parameters as precisely as possible, one should maximize the QFI.
For any given input state $|\psi\rangle_{A}$ with $0 \leq \theta \leq \pi$, and $ 0 \leq \phi < 2\pi$, the optimal QFI with respect to $\theta$ satisfies the necessary and insufficient conditions
\begin{eqnarray}
\left\{\begin{array}{cc}
\frac{\partial\mathcal{F}_{\theta}}{\partial\theta}|_{\theta_{i}}\equiv0, & \mbox{ and } \, (\theta_{i}=0,\frac{\pi}{2},\pi),\nonumber \\
\frac{\partial_{}^{2}\mathcal{F}_{\theta}}{\partial_{}^{2}\theta_{i}}\leq 0, & \mbox{ for } \, (\theta_{i}=0,\frac{\pi}{2},\pi).
\end{array}\right.
\label{eq0}
\end{eqnarray}
We are interested in estimating parameter $\theta_{i}=0,\frac{\pi}{2},\pi$ where the maximum of $\mathcal{F}_{\theta}$ may be located at. (i) for both $\theta=0$ and $\theta=\pi$, one can easily find that $\mathcal{F}_{\theta}|_{\theta=0}=\mathcal{F}_{\theta}|_{\theta=\pi}=\frac{1}{4}(Bc_{1}-Ac_{2})^2\cos^2\phi+\frac{1}{4}(Bc_{2}-Ac_{1})^2\sin^2\phi$, which are related to $\phi$. However, for a class of special initial states' parameters, e.g., $|c_{1}|=|c_{2}|=c$, $\mathcal{F}_{\theta}|_{\theta=0}=\mathcal{F}_{\theta}|_{\theta=\pi}=c^2[1-D(1-\mu)]^2$ is independent of $\phi$. (ii) for $\theta=\frac{\pi}{2}$, $\mathcal{F}_{\theta}|_{\theta=\frac{\pi}{2}}=[c_{3}((2-D)D(1-\mu)-1)-D^2(1-\mu)]^2$. These results are easily determined from Eq.~\eqref{Eq8} where the function $\mathcal{F}_{\theta}$ is symmetric with respect to $\theta=\frac{\pi}{2}$. (iii)based on the conditions $\frac{\partial\mathcal{F}_{\theta}}{\partial\theta}|_{\theta_{i}}\equiv0$ and $\frac{\partial^2\mathcal{F}_{\theta}}{\partial^2\theta}|_{\theta_{i}}>0 (\leq0)$, there exists a threshold value $\mu=\mu^{\star}(c_{i},D,\phi)$ for the estimation degree of parameter located at $\theta_{i}$ with maximum or minimum of $\mathcal{F}_{\theta}$ that not only depends on the initial parameters $c_{i}$ and $\phi$ but also on noisy channel parameters $D$. For $\mu\geq\mu^{\star}(c_{i},D,\phi)$, the maximum of $\mathcal{F}_{\theta}$ locates at $\theta=\frac{\pi}{2}$. While, $\mu<\mu^{\star}(c_{i},D,\phi)$, optimal estimating weight parameter is achieved at $\theta=0$ or $\pi$. Note that $\mu^{\star}(c_{i},D,\phi)$ is determined from equation $\mathcal{F}_{\theta}|_{\theta=0(\pi)}=\mathcal{F}_{\theta}|_{\theta=\frac{\pi}{2}}$. On the other hand, it is worthy pointing out that, if one estimates parameter at $\theta=0$ or $\pi$, corresponding to where the  teleportation state is $|\psi\rangle=|0\rangle$ or $|\psi\rangle=e^{i\phi}|1\rangle$, which only carries classical information. However, if one estimates parameter at $\theta=\pi/2$, corresponding to where the input state (or the  teleportation state) is $|\psi\rangle=\frac{1}{\sqrt{2}}(|0\rangle+e^{i\phi}|1\rangle)$, which carries quantum information.

Similarly, the possible optimal estimation $\phi$ ($\mathcal{F}_{\phi}$) can also be obtained from the conditions
\begin{eqnarray}
\begin{array}{cc}
\frac{\partial\mathcal{F}_{\phi}}{\partial\phi}|_{\phi_{j}}\equiv0, & \mbox{ with } \,  (\phi_{j}=0,\frac{\pi}{2},\pi,\frac{3\pi}{2})\nonumber \\
\end{array}
\label{eq0}
\end{eqnarray}
This implies the estimation parameter located at $\phi_{j}=0,\frac{\pi}{2},\pi,\frac{3\pi}{2}$ where $\mathcal{F}_{\phi}$ has a maximum or minimum value. By inserting $\phi_{j}$ into Eq.~\eqref{Eq9}, one can find that $\mathcal{F}_{\phi}|_{\phi=0}=\mathcal{F}_{\phi}|_{\phi=\pi}=\frac{1}{4}(Bc_{2}-Ac_{1})^2\sin^2\theta$, while $\mathcal{F}_{\phi}|_{\phi=\frac{\pi}{2}}=\mathcal{F}_{\phi}|_{\phi=\frac{3\pi}{2}}=\frac{1}{4}(Bc_{1}-Ac_{2})^2\sin^2\theta$ which are related to $\theta$.
However, for $\theta=0$ or $\pi$, $\mathcal{F}_{\phi}=0$ is independent of $\phi$ and has a minimum value, while for $\theta=\pi/2$, $\mathcal{F}_{\phi}|_{\theta=\pi/2}$ has the optimal parameter estimation. Especially, for a class of special initial states' parameters, e.g., $|c_{1}|=|c_{2}|=c$, $\mathcal{F}_{\phi}=c^2[1-D(1-\mu)]^2$.

\subsection{Phase-damping channel with memory}
Phase damping channel which is an unital channel describes a quantum noise with loss of quantum phase information but not loss of energy. The Kraus operators for quantum dephasing channel with memory are given in Table I, where $P_{0}=1-D$, $P_{3}=D$. Submitting Eq.~\eqref{eq3} into ~\eqref{eq7}, one gets the final teleported state whose three Bloch vector components are
\begin{eqnarray}
r_{x}&=&\Delta c_{1}\cos\phi\sin\theta\nonumber \\
r_{y}&=&\Delta c_{2}\sin\phi\sin\theta\nonumber \\
r_{z}&=&-c_{3}\cos\theta
\end{eqnarray}
with $\Delta=[1-(2-D)D(1-\mu)]$.
Using Eq.~\eqref{eq1}, the teleported QFI with respect to $\theta$ is
\begin{eqnarray}\label{Eq10}
\mathcal{F}_{\theta}&=&\Delta^{2}\cos^{2}\theta(c_{1}^{2}\cos^{2}\phi+c_{2}^{2}\sin^{2}\phi)+c_{3}^2\sin^{2}\theta\nonumber\\
&+&\frac{\cos^2\theta\sin^2\theta[\Delta^{2}(c_{1}^{2}+c_{2}^{2})\cos^{2}\phi-c_{3}^2]^2}{1-c_{3}^2\cos^{2}\theta-\Delta^{2}\sin^{2}\theta(c_{1}^{2}\cos^{2}\phi+c_{2}^{2}\sin^{2}\phi)}
\end{eqnarray}
which also depends on the initial parameters $c_{i}$, input sate parameters $\theta$ and $\phi$, se well as noisy channel parameters $D$ and $\mu$. To acquire a high parameter estimation precision, similar to the previous analysis, $\mathcal{F}_{\theta}$ has extremum values satisfying the conditions
$\frac{\partial\mathcal{F}_{\theta}}{\partial\theta}\equiv0$ with respect to $\theta_{i}=0,\frac{\pi}{2},\pi$. When estimated $\theta=\frac{\pi}{2}$, $\mathcal{F}_{\theta}|_{\theta=\frac{\pi}{2}}=c_{3}^{2}$ is unaffected by phase-damping channel. This implies that the information encoded in input state with $\theta=\frac{\pi}{2}$ is immune to the phase-damping channel. On the other hand, when estimated $\theta=0$ or $\pi$, we have $\mathcal{F}_{\theta}|_{\theta=0}=\mathcal{F}_{\theta}|_{\theta=\pi}=\Delta^2(c_{1}^{2}\cos^2\phi+c_{2}^{2}\sin^2\phi)$ which are related to $\phi$ and affected by the phase-damping noise. Like the case of the amplitude damping channel with memory, there exists a threshold value $\mu^{\star}=\frac{\sqrt{2}|c_{3}|-(1-D)^2\sqrt{c_{1}^{2}+c_{2}^{2}+(c_{1}^{2}-c_{2}^{2})\cos2\phi}}{(2-D)D\sqrt{c_{1}^{2}+c_{2}^{2}+(c_{1}^{2}-c_{2}^{2})\cos2\phi}}$ in the degree of memory above which $\mathcal{F}_{\theta}$ has optimal parameter estimation precision at $\theta=\frac{\pi}{2}$.
Below the threshold $\mathcal{F}_{\theta}$ has optimal parameter estimation precision at $\theta=0$ or $\pi$. In other words, the teleported state with $\theta=\pi/2$, corresponding to $|\psi\rangle=\frac{1}{\sqrt{2}}(|0\rangle+e^{i\phi}|1\rangle)$ carrying quantum information is higher parameter estimation precision than the one $\theta=0$ or $\pi$, corresponding to $|\psi\rangle=|0\rangle$ or $|\psi\rangle=e^{i\phi}|1\rangle$ carrying classical information strongly depends on the threshold value.

Next, we analytically evaluate QFI with respect to $\phi$ under the phase-damping noise. In this
case, $\mathcal{F}_{\phi}$ is given by
\begin{eqnarray}\label{Eq11}
\mathcal{F}_{\phi}&=&\Delta^{2}\sin^{2}\theta(c_{1}^{2}\sin^{2}\phi+c_{2}^{2}\cos^{2}\phi)\nonumber\\
&+&\frac{[\Delta^{2}(c_{1}^{2}-c_{2}^{2})\sin^{2}\theta\cos\phi\sin\phi]^2}{1-c_{3}^2\cos^{2}\theta-\Delta^{2}\sin^{2}\theta(c_{1}^{2}\cos^{2}\phi+c_{2}^{2}\sin^{2}\phi)}
\end{eqnarray}

In a similar fashion, we focus on following cases: $\mathcal{F}_{\phi}|_{\phi=0}=\mathcal{F}_{\phi}|_{\phi=\pi}=\Delta^2c_{2}^{2}\sin^2\theta$, and $\mathcal{F}_{\phi}|_{\phi=\frac{\pi}{2}}=\mathcal{F}_{\phi}|_{\phi=\frac{3\pi}{2}}=\Delta^2c_{1}^{2}\sin^2\theta$.
However, for $\theta=0,\pi$, $\mathcal{F}_{\phi}|_{\theta=0}=\mathcal{F}_{\phi}|_{\theta=\pi}=0$ regardless of $\phi$ has a minimum value,  while for $\theta=\pi/2$, $\mathcal{F}_{\phi}|_{\theta=\pi/2}$ has the optimal parameter estimation. Especially, for a class of special initial states' parameters, e.g., $|c_{1}|=|c_{2}|=c$, $\mathcal{F}_{\phi}=c^2[1-(2-D)D(1-\mu)]^2$ is independent of $\phi$ and has a maximum value.

\subsection{Depolarizing channel with memory}
Depolarizing channel is a Pauli channel with particularly nice symmetry properties and the Kraus operators in the presence of memory are given in Table I, where $P_{0}=1-D$, $P_{1}=P_{2}=P_{3}=D/3$. According to Eqs.~\eqref{eq3} and~\eqref{eq7}, the corresponding three Bloch vector components of the teleported state are
\begin{eqnarray}
r_{x}&=&\frac{1}{9}\Lambda c_{1}\cos\phi\sin\theta\nonumber \\
r_{y}&=&\frac{1}{9}\Lambda c_{2}\sin\phi\sin\theta\nonumber \\
r_{z}&=&\frac{1}{9}\Lambda c_{3}\cos\theta
\end{eqnarray}
with $\Lambda=[9-8(3-2D)D(1-\mu)]$.
Similarly, the analytic expression of QFI with respect to $\theta$ is obtained
\begin{eqnarray}\label{Eq12}
\mathcal{F}_{\theta}&=&\frac{1}{81}\{\Lambda^{2}( c_{1}^{2}\cos^2\theta\cos^{2}\phi+ c_{2}^{2}\cos^2\theta\sin^{2}\phi+c_{3}^2\sin^{2}\theta)\nonumber\\
&-&\frac{\sin^2\theta\cos^2\theta\Lambda^{4}[(c_{1}^{2}-c_{3}^{2})-\sin^2\phi(c_{1}^{2}-c_{2}^2)]^2}{\Lambda^{2}(c_{1}^{2}\sin^2\theta\cos^{2}\phi + c_{2}^{2}\sin^2\theta\sin^{2}\phi+ c_{3}^2\cos^{2}\theta)-81}\}\nonumber\\
\end{eqnarray}

A similar analysis is valid for depolarizing channel, $\mathcal{F}_{\theta}$ has the local maximal or minimal value, if only if
$\frac{\partial\mathcal{F}_{\theta}}{\partial\theta}\equiv0$. For $\theta=\frac{\pi}{2}$, Eq.~\eqref{Eq12} reduces to
$\mathcal{F}_{\theta}|_{\theta=\frac{\pi}{2}}=\frac{1}{81}\Lambda^2 c_{3}^{2}$ is not depended on $\phi$, while for both $\theta=0$ and $\pi$, Eq.~\eqref{Eq12} reduces to
$\mathcal{F}_{\theta}|_{\theta=0}=\mathcal{F}_{\theta}|_{\theta=\pi}=\frac{1}{81}\Lambda^2(c_{1}^{2}\cos^2\phi+c_{2}^{2}\sin^2\phi)$. However, different from the above decoherence channels, $\mathcal{F}_{\theta}$ has optimal parameter estimation precision at $\theta=\frac{\pi}{2}$ if only if $c_{3}^{2}\geq (c_{1}^{2}\cos^2\phi+c_{2}^{2}\sin^2\phi)$, otherwise, $\mathcal{F}_{\theta}$ has optimal parameter estimation precision at $\theta=0$ or $\pi$.

Following, $\mathcal{F}_{\phi}$ suffering from depolarizing channel can be calculated
\begin{eqnarray}\label{Eq13}
\mathcal{F}_{\phi}&=&\frac{1}{81}\{\Lambda^{2}\sin^2\theta( c_{1}^{2}\sin^{2}\phi+c_{2}^{2}\cos^2\phi)\nonumber\\
&-&\frac{(c_{1}^{2}-c_{2}^{2})^2\Lambda^{4}\sin^4\theta\sin^2\phi\cos^2\phi}{\Lambda^{2}( c_{1}^{2}\sin^2\theta\cos^{2}\phi+c_{2}^{2}\sin^2\theta\sin^{2}\phi + c_{3}^2\cos^{2}\theta)-81}\}\nonumber\\
\end{eqnarray}
As we expect, $\frac{\partial\mathcal{F}_{\phi}}{\partial\phi}|_{\phi_{j}}\equiv0$, with $\phi_{j}=0,\frac{\pi}{2},\pi,\frac{3\pi}{2}$, and $\mathcal{F}_{\phi}|_{\phi=0}=\mathcal{F}_{\phi}|_{\phi=\pi}$, while $\mathcal{F}_{\phi}|_{\phi=\frac{\pi}{2}}=\mathcal{F}_{\phi}|_{\phi=\frac{3\pi}{2}}$ is dependent on $\theta$. Especially, for $\theta=\frac{\pi}{2}$,
$\mathcal{F}_{\phi}|_{\phi=0}=\mathcal{F}_{\phi}|_{\phi=\pi}=\frac{1}{81}\Lambda^2c_{2}^{2}$, and
$\mathcal{F}_{\phi}|_{\phi=\frac{\pi}{2}}=\mathcal{F}_{\phi}|_{\phi=\frac{3\pi}{2}}=\frac{1}{81}\Lambda^2c_{1}^{2}$. However, for $\theta=0$ or $\pi$, $\mathcal{F}_{\phi}=0$. Besides, for a class initial states as resources satisfied condition $|c_{1}|=|c_{2}|=c$, $\mathcal{F}_{\phi}|_{\theta=\pi/2}=\frac{1}{81}c^2[9-8(3-2D)D(1-\mu)]^2$  has a maximum value regardless of $\phi$.
\begin{figure}[htpb]
   \includegraphics[width=4.5cm]{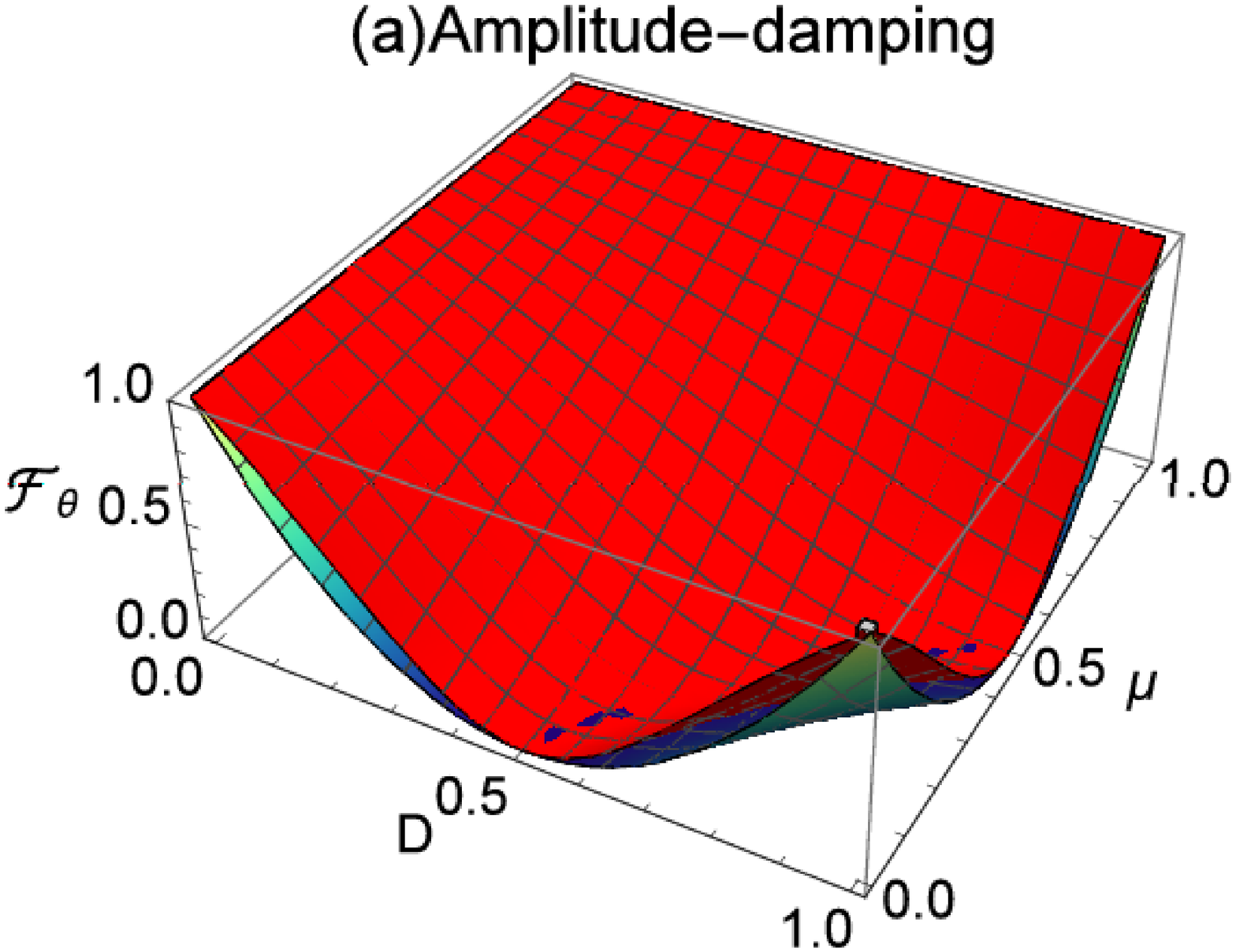}
   \includegraphics[width=4.5cm]{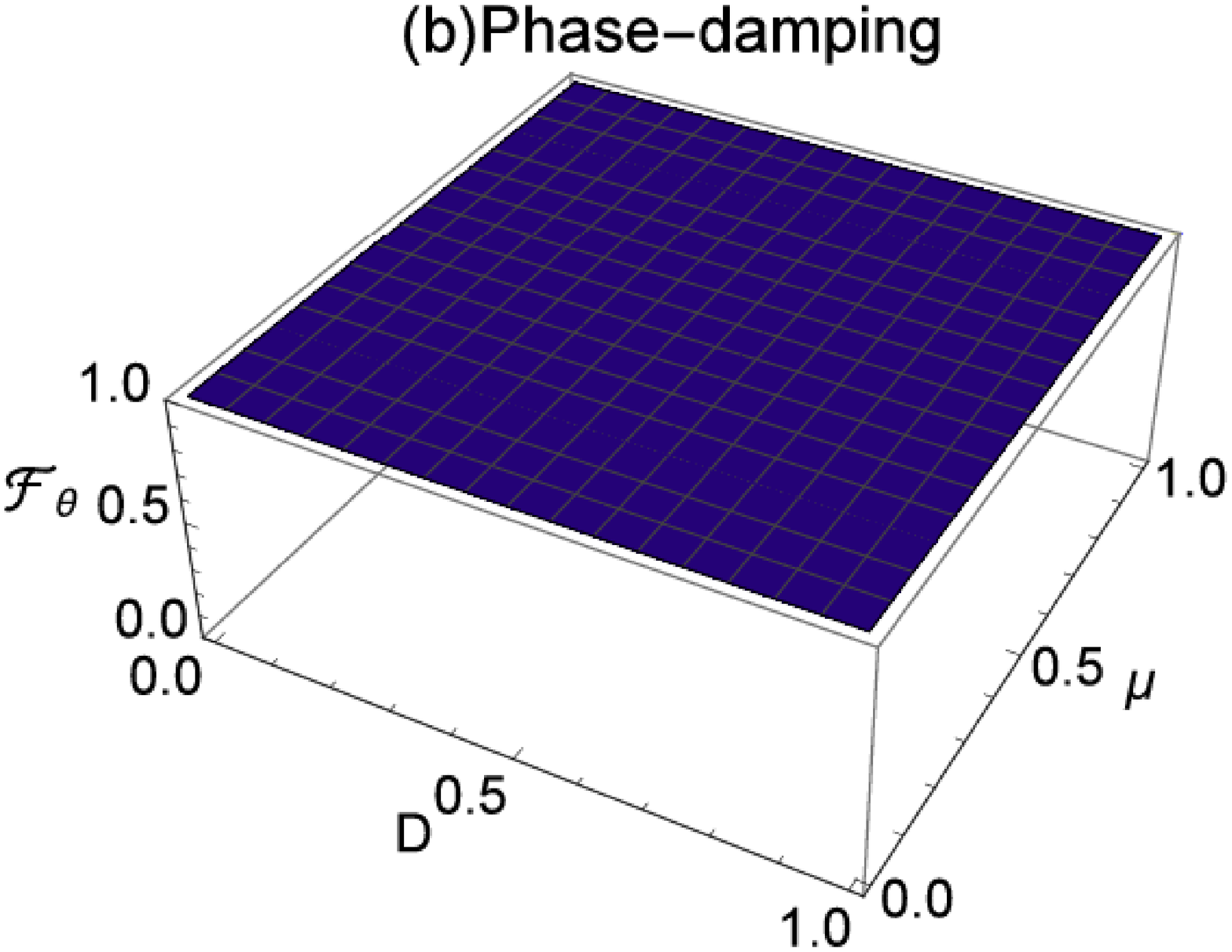}
   \includegraphics[width=4.5cm]{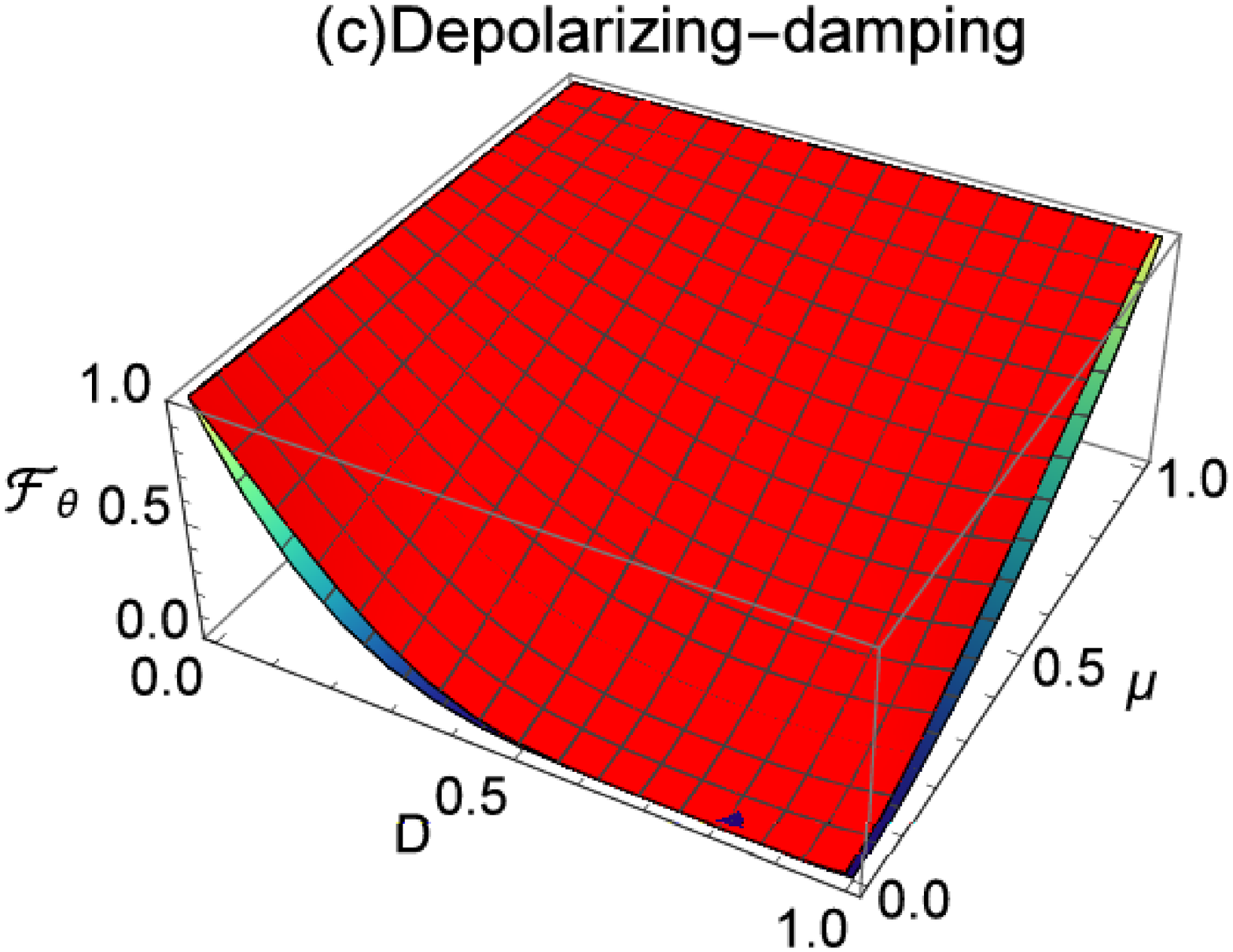}
 \caption{\label{Fig1}(Color online) QFI with respect to $\theta$ (the color
plot corresponding to $|\psi\rangle_{A}$, the red plot corresponding to$|\psi\rangle_{AB}$) under noisy channels with memory as functions of $D$ and $\mu$ for given parameters $\theta=\pi/2$, $c_{1}=c_{2}=\pm1$ and $c_{3}=-1$.}
\end{figure}

\section{Entanglement teleportation}

Next we extend to the teleportation of QFI for a two-qubit case in the presence of memory effect, and focus on how the decoherence channels with memory affect teleportation of information of specific parameters encoded in quantum state. It has been proved theoretically that, teleportation of an unknown two-qubit state through two independent and equally entangled, noisy quantum channels is equivalent to a generalized depolarizing channel
with probabilities given by the maximally entangled components of the resource by Lee and Kim\cite{Lee}. In fact, their protocol can be performed by doubling protocol $\mathcal{T}_{0}$. Therefore, we use two copies of the two-qubit state $\varepsilon(\rho)$ shared between the sender and receiver as quantum channel to teleport another two-qubit state $|\psi\rangle_{AB}$, namely,
\begin{eqnarray}
\Lambda(|\psi\rangle_{AB}\langle\psi|)&=&\sum_{i,j=0}^{3}\langle\Psi_{Bell}^{i}|\varepsilon(\rho)|\Psi_{Bell}^{i}\rangle\langle\Psi_{Bell}^{i}|\varepsilon(\rho)|\Psi_{Bell}^{i}\rangle\nonumber\\
&\times&(\sigma_{i}\otimes \sigma_{j})|\psi\rangle_{AB}\langle\psi|(\sigma_{i}\otimes \sigma_{i})
\label{eq70}
\end{eqnarray}
We here, consider a two-qubit
entangled state $|\psi\rangle_{AB}=\cos\frac{\theta}{2}|00\rangle+e^{i\phi}\sin\frac{\theta}{2}|11\rangle$
with $(0 \leq \theta \leq \pi, 0 \leq \phi < 2\pi)$ as the teleported state in the process of teleportation. It is not difficult found that, the QFI of two qubits state $|\psi\rangle_{AB}$ with respect to $\theta$ and $\phi$ before teleportation, is equivalent to those of single qubit state $|\psi\rangle_{A}$: $\mathcal{F}_{\theta}=1$ and $\mathcal{F}_{\phi}=\sin^2\theta$, respectively. Using Eq.~\eqref{eq70}, the QFIs of the teleported state can be obtained exactly by Eq~\eqref{eq2}. In the following, we carry out a numerical study the initial state $\rho=|\Psi_{}^{\pm}\rangle\langle\Psi_{}^{\pm}|$, with $|\Psi_{}^{\pm}\rangle=\frac{1}{\sqrt{2}}(|01\rangle\pm|10\rangle)$ as resources, namely, $c_{1}=c_{2}=\pm1$ and $c_{3}=-1$. We are interested to investigate the precision of estimating the parameters $\theta$ and $\phi$ which are encoded in the quantum resources teleportation state.
\begin{figure}[htpb]
   \includegraphics[width=4.5cm]{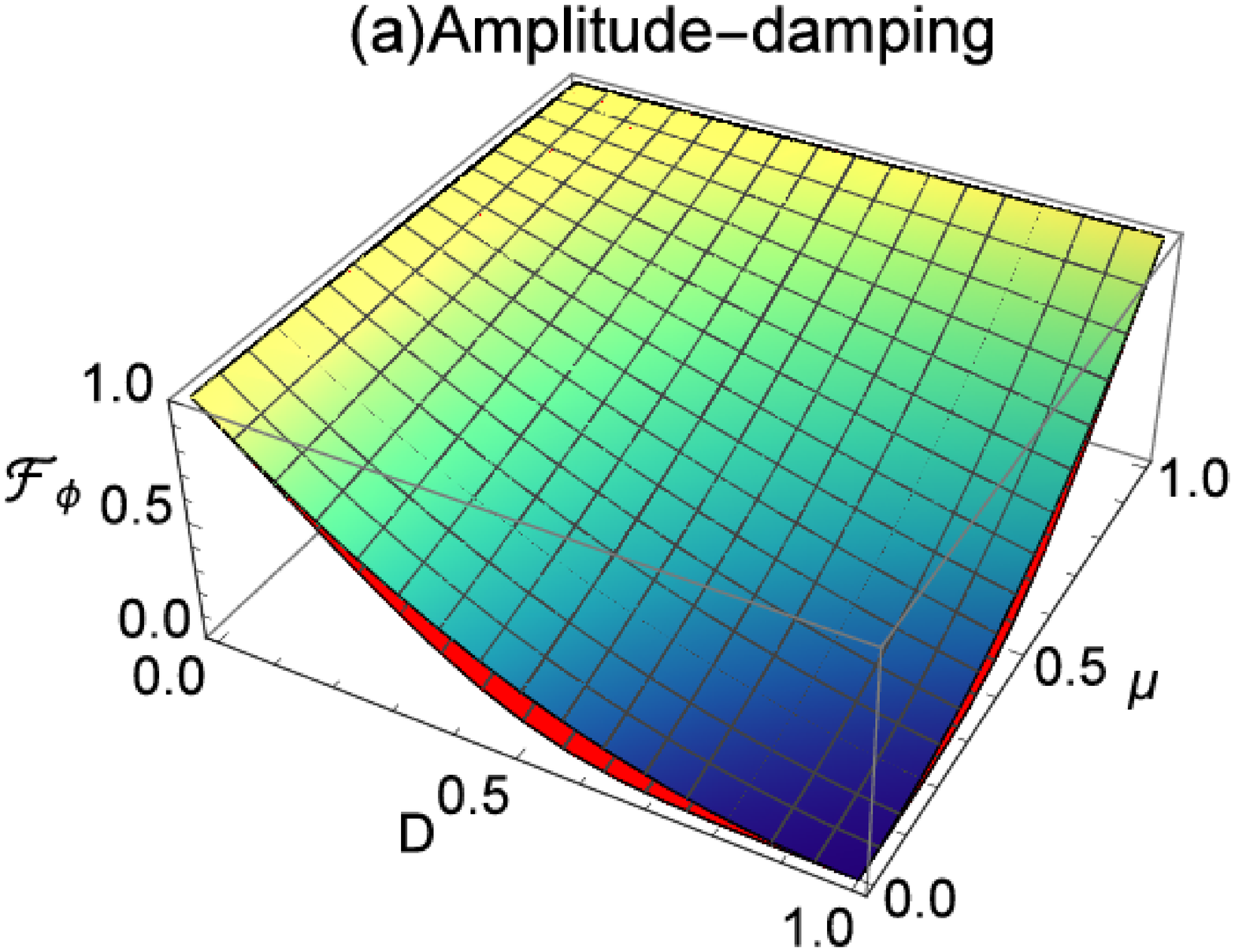}
   \includegraphics[width=4.5cm]{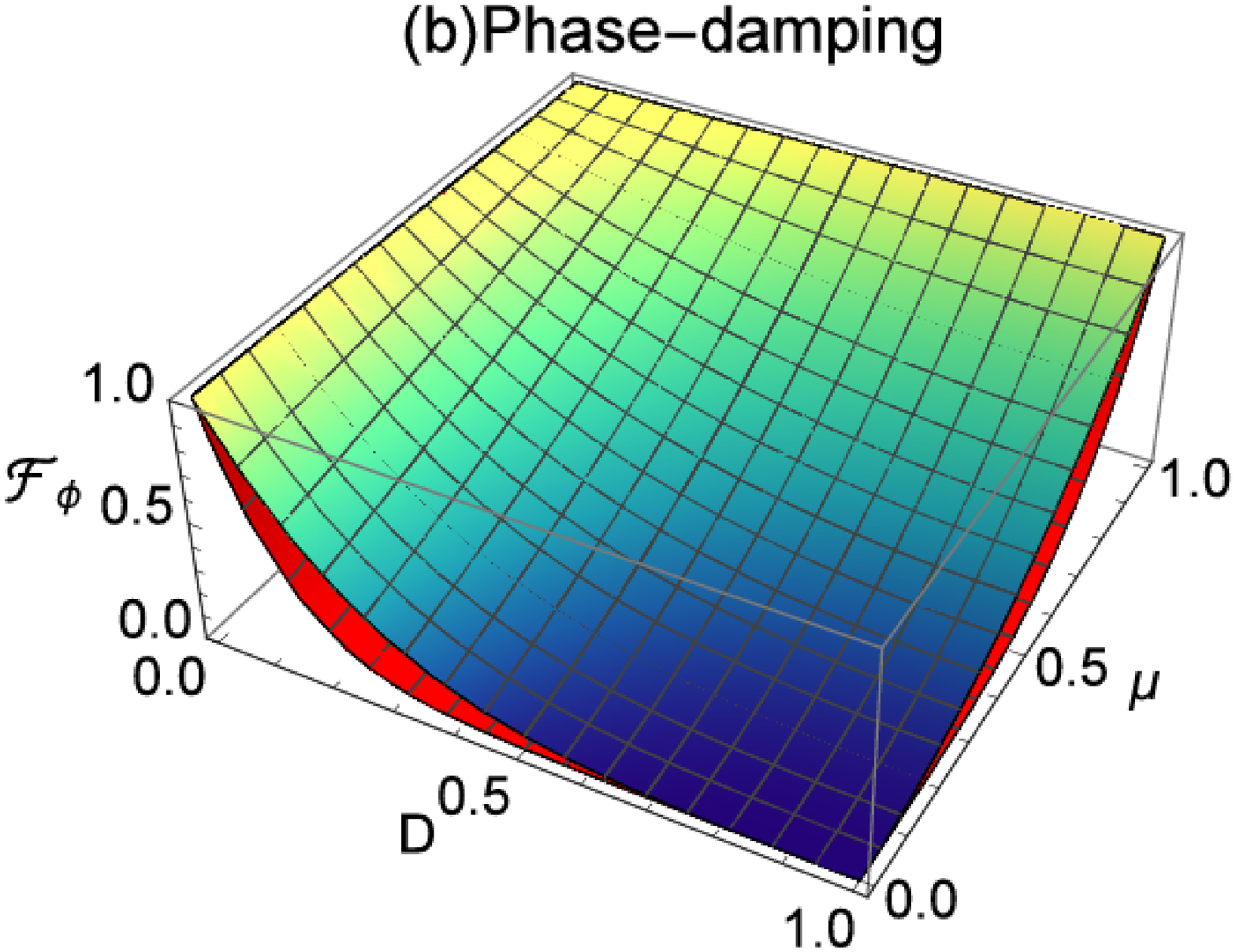}
   \includegraphics[width=4.5cm]{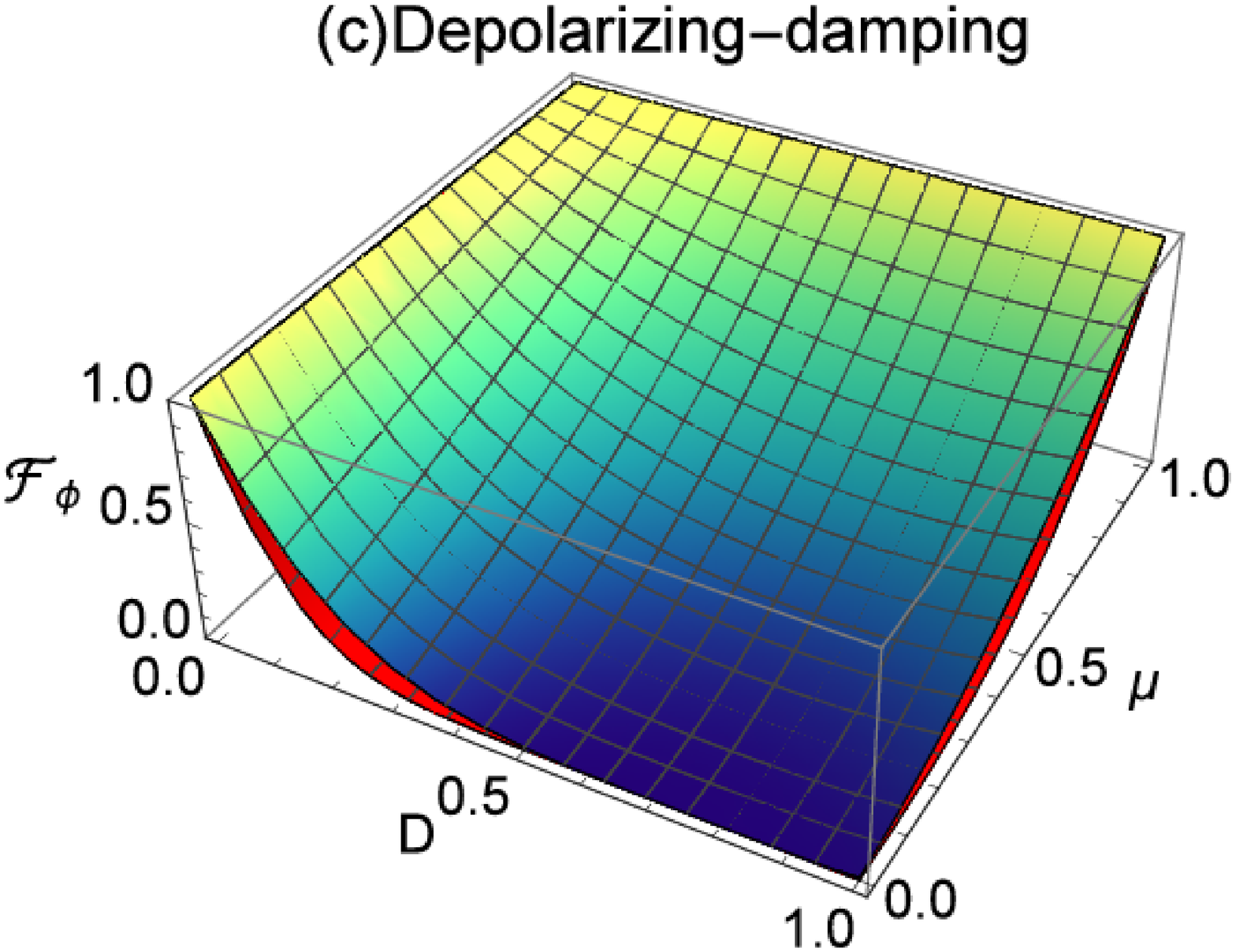}
 \caption{\label{Fig2}(Color online) QFI with respect to $\phi$ (the color
plot corresponding to $|\psi\rangle_{A}$, the red plot corresponding to$|\psi\rangle_{AB}$) under noisy channels with memory as functions of $D$ and $\mu$ for given parameters $\theta=\pi/2$, $c_{1}=c_{2}=\pm1$ and $c_{3}=-1$.}
\end{figure}

In order to make results comparable, we have plotted
the results of QFI teleportation for both single- and two-qubit case in both Figs.(1) and (2).
It is clearly shown that the QFI teleportation of $|\psi\rangle_{A}$ and $|\psi\rangle_{AB}$ display the same behaviors. Fig.(1) is shown the QFI for teleported states  $|\psi\rangle_{A}$ and $|\psi\rangle_{AB}$ with respect to $\theta$ as functions of $D$ and $\mu$ under different decoherence channels. From this graph, several distinct features can be found with the increasing $D$ and $\mu$. First, the QFI has the minimal value at $D=\frac{c_{3}}{1+c_{3}}$ for $0\leq c_{3}\leq 1$ or at $D=\frac{c_{3}}{c_{3}(1-\mu)-\sqrt{(c_{3}+c_{3}^{2}\mu)(\mu-1)}}$ for $-1\leq c_{3}\leq 0$ under amplitude damping channel. Second, the QFI is constant and not affected by the phase damping channel noise. Third, the behavior of QFI monotonously decays with $D$ increasing but progressively increases with $\mu$ increasing under the depolarizing channel. Finally, the amount of QFI teleportation of $|\psi\rangle_{A}$ is larger than or equal to that of $|\psi\rangle_{AB}$ with respect to $\theta$. On the other hand, Fig.(2) is shown the QFI for teleported states  $|\psi\rangle_{A}$ and $|\psi\rangle_{AB}$ with respect to $\phi$ as functions of $D$ and $\mu$ under different decoherence channels. As we can see, the same behaviors of QFI display. At the same time, increasing the degree of channels memory $\mu$ leads to improve the precision of estimating phase parameter $\phi$. However, contrary to the precision of estimating weight parameter, the amount of QFI teleportation of $|\psi\rangle_{A}$ with respect to $\phi$ is smaller than that of $|\psi\rangle_{AB}$.

\section{Conclusion}  %%% ½Ú±êÌâ 1

In summary, we have explored the teleportation of QFI for a class
of X-states as resources under quantum channels with memory. We concern only the transformation of information of specifc parameter of
the quantum state. Resort to the QFI definition, the explicit analytical results for the estimation precision of an unknown parameters $\theta$ and $\phi$ under three different decoherence channels are obtained. We find that the precision of estimating weight parameter $\theta$ and the precision of estimating phase
parameter $\phi$, are determined by the initial parameters $c_{i}$, input sate parameters $\theta$ and $\phi$, as well as noisy channel parameters $D$ and $\mu$ during the teleportation process. Meanwhile, the teleportation of a two-qubit system has also been considered. Taking Bell entanglement states as resources as example, we compare both the situations where the teleportation of QFI for single-and two-qubit states to teleport. The remarkable similarities and differences among these two situations under noisy channels are also analyzed in detail and some significant results are presented. We show that memory effects can be improved the precision of estimating parameters.

We should finally mention that even though we have examined a particular X-state as resource, which shares between partners under the consecutive uses of quantum channels,
and also considered a specific type of Bell entanglement teleportation, our treatment can be easily applied to study more
general scenarios in a straightforward way.
\acknowledgments
This work is supported by the National Basic Research Program of China under Grant No. 2016YFA0301903, and the National Natural Science Foundation of China under Grants Nos. 11747107, 11174370, 11304387, 61632021, 11305262, 61205108 and 11574398, the Natural Science Foundation of Hunan Province (Grant No.2017JJ3346), the Project of Science and Technology Plan of Changsha (K1705022 and kc1809023), and Key Laboratory of Low-Dimensional Quantum Structures and Quantum Control of Ministry of Education (QSQC1810)

\appendix

\section{}
\label{app:eff-trans}
In this appendix, we give the Kraus operators for amplitude damping(Am), phase damping(Pd), and depolarizing(De) channels with memory:

\begin{widetext}
\makeatletter\def\@captype{table}\makeatother
\caption{Kraus operators for amplitude damping(Am), phase damping(Pd), and depolarizing(De) channels with memory, where $D$ represents the decoherence parameter.}$
\begin{tabular}{|c|c|c|c|}
\hline
$\text{Channel description}$&$\text{ Uncorrelated Kraus operators}$&$\text{Correlated Kraus operators}$\\
\hline
$\text{Am}$ & $
\begin{tabular}{l} $E_{ij}^{u}=A_{i}\otimes A_{j},\quad (i,j=0,1)$ \\ $A_{0}=\left[
\begin{array}{cc}
\sqrt{1-D} & 0 \\
0 & 1%
\end{array}%
\right] ,$ $A_{1}=\left[
\begin{array}{cc}
0 & 0 \\
\sqrt{D} & 0%
\end{array}%
\right] $
\end{tabular}
$ & $
E_{00}^{c}=\left[
\begin{array}{cccc}
\sqrt{1-D} & 0 & 0 & 0 \\
0 & 1 & 0 & 0 \\
0 & 0 & 1 & 0 \\
0 & 0 & 0 & 1
\end{array}%
\right] ,$ $E_{11}^{c}=\left[
\begin{array}{cccc}
0 & 0 & 0 & 0 \\
0 & 0 & 0 & 0 \\
0 & 0 & 0 & 0 \\
\sqrt{D} & 0 & 0 & 0
\end{array}%
\right]
$ \\
\hline
$\text{Pd}$ &  $
\begin{tabular}{l} $E_{ij}^{u}=\sqrt{P_{i}P_{j}}\sigma_{i}\otimes \sigma_{j},\quad (i,j=0,3)$ \\ $\sigma_{0}=\left[
\begin{array}{cc}
1 & 0 \\
0 & 1%
\end{array}%
\right] ,$ $\sigma_{3}=\left[
\begin{array}{cc}
1 & 0 \\
0 & -1%
\end{array}%
\right] $
\end{tabular}
$ &  $
\begin{tabular}{l} $E_{kk}^{c}=\sqrt{P_{k}}\sigma_{k}\otimes \sigma_{k},\quad (k=0,3)$ \\ $\sigma_{0}=\left[
\begin{array}{cc}
1 & 0 \\
0 & 1%
\end{array}%
\right] ,$ $\sigma_{3}=\left[
\begin{array}{cc}
1 & 0 \\
0 & -1%
\end{array}%
\right] $
\end{tabular}
$ \\
\hline
$\text{De}$ & $
\begin{tabular}{l} $E_{ij}^{u}=\sqrt{P_{i}P_{j}}\sigma_{i}\otimes \sigma_{j},\quad (i,j=0,1,2,3)$ \\ $\sigma_{0}=\left[
\begin{array}{cc}
1 & 0 \\
0 & 1%
\end{array}%
\right] ,$ $\sigma_{1}=\left[
\begin{array}{cc}
0 & 1 \\
1 & 0%
\end{array}%
\right] $ \\ $\sigma_{2}=\left[
\begin{array}{cc}
0 & -i \\
i & 0%
\end{array}%
\right] ,$ $\sigma_{3}=\left[
\begin{array}{cc}
1 & 0 \\
0 & -1%
\end{array}%
\right] $
\end{tabular}
$ &  $
\begin{tabular}{l} $E_{kk}^{c}=\sqrt{P_{k}}\sigma_{k}\otimes \sigma_{k},\quad (k=0,1,2,3)$ \\ $\sigma_{0}=\left[
\begin{array}{cc}
1 & 0 \\
0 & 1%
\end{array}%
\right] ,$ $\sigma_{1}=\left[
\begin{array}{cc}
0 & 1 \\
1 & 0%
\end{array}%
\right] $\\ $\sigma_{2}=\left[
\begin{array}{cc}
0 & -i \\
i & 0%
\end{array}%
\right] ,$ $\sigma_{3}=\left[
\begin{array}{cc}
1& 0 \\
0 & -1%
\end{array}%
\right] $
\end{tabular}
$ \\
\hline
\end{tabular}%
$%
\end{widetext}

\label{app:eff-trans}

\end{document}